# 1/f noise from the coincidences of similar single-sided random telegraph signals

Giovanni Zanella

*Dipartimento di Fisica dell'Università di Padova and Istituto Nazionale di Fisica Nucleare, Sezione di Padova, via  Marzolo 8, 35131 Padova, Italy*

In this paper it is demonstrated that *1/f* power spectrum appears in the process originated by the superposition of many single-sided random telegraph signals (RTS or RTN) with the same amplitude, probability and relaxation time. Indeed, the coincidences of these RTSs generate self-organizing fluctuations which are responsible for the generation of *1/f* noise under various aspects. The accord of the exposed model with other models and with various experimental results is displayed.

## 1. Introduction

  The main feature of 1/f noise is that its power spectrum increases with decreasing frequency $f$ down to the lowest possible frequencies for conducting measurements. This noise is therefore spectrally *scaling*, that is, it is statistically identical to its transformation by contraction in time or another independent variable, followed by a corresponding change in intensity. This scaling property is typical of the *fractals* [1] and it reveals the presence of long-term correlations.

The self-similar (fractal) structure of 1/f noise is revealed by the constancy of its average power per decade and by the scale invariance of its autocorrelation function $\psi(\theta)$. Indeed, if $S(f) = 1/f$, tanks to the Wiener-Khintchine theorem, we have

$$\psi(\theta) = \mathfrak{I}^{-1}\left[\frac{1}{f}\right] = \mathfrak{I}^{-1}\left[\left(\frac{1}{\alpha}\right)\left(\frac{\alpha}{f}\right)\right] = \psi(\alpha\theta) , \qquad (1)$$

being $\mathfrak{I}^{-1}$ the inverse Fourier transform. Hence, 1/f noise cannot be characterized by a single relaxation time $\tau$. In fact,  if $\psi(\theta) = \sigma^2 \exp(-|\theta|/\tau)$, then $\psi(\alpha\theta) \neq \psi(\theta)$, being $\sigma^2$ the variance of the fluctuation.

 Despite great progress being made in *1/f* noise physics, the source of these fluctuations remains unknown in the case of most systems and the problem remains largely unsolved in its generality.

 It would be too lengthy to report all references concerning *1/f* noise but an extensive bibliography has been drawn up in [2] and [3].

 The present work furnishes a further contribution, and important settlements, to a model introduced in the references [4] [5]. This model explains *1/f* noise as due to the superposition of many similar bistable fluctuators having the same power spectrum, in other words by the superposition of many single-sided random telegraph signals (RTS), or random telegraph noises (RTN), with same amplitude, probability and relaxation time $\tau$. These two-state fluctuations can be generated by elementary physical processes such as the intermittent presence of same objects, or physical quantities, on microscopic or macroscopic scale.

Commonly, $1/f$ noise is explained by a weighed summation of many similar independent Lorentzians with a broad distribution of relaxation times, without a general characterisation and justification [2][6]. Indeed, "*the question of the independence of the processes is hardly ever discussed when this is the essential problem*" [7].

We shall see that the analysis of the *coincidences* (the number $c$ of contemporary up levels during the sampling time $\Delta T$) of many similar single-sided RTSs demonstrates a more general origin of $1/f$ noise, where the possible superposition of independent Lorentzians, with a distribution of relaxation times, appears as consequence and not as cause.

In the following, we refer only to time phenomena, approaching gradually the problem.

## 2. Superposition of similar signals

In general, if $U(t)$ is a *stationary process* its autocorrelation function $\psi(\theta)_U$ must be independent from the time $t$, while $\psi(\theta)_U$ depends from the shift time $\theta$.

The autocorrelation function of $U(t)$ is given by

$$\psi(\theta)_U = \lim_{T \to \infty} \frac{1}{T} \int_{-T/2}^{T/2} U(t) \, U(t - \theta) \, dt \quad , \qquad (2)$$

where the integration interval, between $-T/2$ and $T/2$, can be chosen at any point. In practice, the *limit* of the interval $T$ corresponds to the *measurement time*.

In general, the process $U$, resulting from the summation of $N$ signals $u_i$ has the following autocorrelation function

$$\psi(\theta)_U = \lim_{T \to \infty} \frac{1}{T} \int_{-T/2}^{T/2} \left[ \sum_{i=1}^{N} u_i(t) \right] \left[ \sum_{j=1}^{N} u_j(t - \theta) \right] dt \quad . \qquad (3)$$

Thus, we can write

$$\psi(\theta)_U = \lim_{T \to \infty} \frac{1}{T} \sum_{i=1}^{N} \sum_{j=1}^{N} \int_{-T/2}^{T/2} u_i(t) \, u_j(t - \theta) \, dt \quad , \qquad (4)$$

where the terms with $i=j$ denote autocorrelation functions, while the terms with $i \neq j$ are of cross-correlation.

Operating in *stationary regime* (called *relaxed* in ref. [5]), the sum of $N$ similar stationary and stochastic signals $u_i$ furnishes identical autocorrelation functions $\psi_a(\theta)$, for any $i=j$, and identical cross-correlation functions $\psi_c(\theta)$, for any $i \neq j$. Hence, the autocorrelation function of the resulting process $U$, will be

$$\psi(\theta)_U = N\psi_a(\theta) + N(N-1)\psi_c(\theta) \quad . \tag{5}$$

Averagely, due the randomness of the signals

$$\psi_c(\theta) = \langle u_i u_j \rangle = \langle u_i \rangle^2 \qquad \text{(for } i \neq j) \ . \tag{6}$$

It is important to note that Eqs.(5) and (6), concern stationary and stochastic signals. In the case of originating single-sided RTSs, the term of cross-correlation is just due to the random presence of coincidences. Indeed, if hypothetically the coincidences are missing, thus $\psi_c(\theta) = 0$.

The Wiener-Khintchine theorem (expressed for stationary and stochastic processes [8]) permits us to find the power spectrum $S(\omega)_U$ of $U$, by the Fourier transform of $\psi(\theta)_U$. So, from Eq.(5) we have

$$S(\omega)_U = N\,\Im[\psi_a(\theta)] + N(N-1)\,\Im[\psi_c(\theta)] \ , \tag{7}$$

where $\Im[\psi_a(\theta)] = S(\omega)_u$ and $\Im[\psi_c(\theta)]$ are the Fourier transform of $\psi_a(\theta)$ and $\psi_c(\theta)$.

Therefore, Eq.(7) may be written in the form

$$S(\omega)_U = NS(\omega)_u + N(N-1)\langle u_i \rangle^2 \delta(\omega) \quad , \tag{8}$$

where $\delta(\omega)$ is the Dirac-delta.

In conclusion, in the sum of $N$ signals which have the same power spectrum $S(\omega)_u$, the resulting power spectrum $S(\omega)_U$ consists of two additive terms: the term $N\,S(\omega)_u$ independent from cross-correlations, and the cross-correlation term, which is due to the *coincidences* in case of single-sided originating RTSs.

The power spectrum $S(\omega)_U$ represents also the power spectrum $S(\omega)_{\Delta U}$ of the fluctuations $\Delta U$ from the mean $<U>$, a part a dc component

## 3. Superposition of single-sided rectangular signals

It is useful to consider the superposition of $N \gg 1$, random-phased, single-sided rectangular signals $u_i$ of the same: amplitude $u$, pulse duration and period $\tau_0$.

The process obtained summing these signals consists of various trains of pulses of different amplitude and shape, every with period $\tau_0$ (see the example of Fig.1 for five signals).

A train of pulses can emerge upon the others, in the resulting process, when it is due to a number of coincidences (in up level) which exceeds those of other trains. In any case this emersion is guaranteed if $c > N/2$.

In particular, the whole coincidence of all the originating signals produces a single rectangular process of amplitude $Nu$ and period $\tau_0$. In this case the autocorrelation function will be

$$\psi(\theta)_U = N\psi_a(\theta) + N(N-1)\psi_a(\theta) = N^2\,\psi_a(\theta)\quad. \tag{9}$$

The lack of coincidences (in up level) would yield instead $\psi(\theta)_U = N\psi_a(\theta)$.

Therefore, the term of cross-correlation involves only coincidences and it can range from zero to $N(N-1)\psi_a(\theta)$, depending on the phasing of the originating signals. In case of random phasing of the originating signals the cross-correlation term would be statistically steady, that this $N(N-1) <u_i>^2$.

As far as it is concerned the memory of a periodic signal, being the value of the signal at a done instant correlated to the value of the signal at every other instant, the memory is infinite.

## 4. Statistical properties of an RTS

Before to consider the superposition of many similar single-sided RTSs it is useful to recall the statistical properties of a single RTS [9].

An RTS has only one relaxation time $\tau$ and this corresponds to the relationship

$$\tau = 2\frac{\tau_u \tau_d}{\tau_u + \tau_d}\quad. \tag{10}$$

where $\tau_u$ denotes the mean lifetime in up level and $\tau_d$ that in down level.

The consequences of Eq.(10) are important. Indeed, as $\tau$ is accepted to represent the memory of the signal, the maximum of memory corresponds to $\tau_u = \tau_d = \tau$, instead if $\tau_u \ll \tau_d$, or vice versa, $\tau \cong \tau_u$, or $\tau \cong \tau_d$. In this last cases the RTS approaches a Poisson process.

The probability $p_u$ of up level is

$$p_u = \frac{\dfrac{1}{\tau_d}}{\dfrac{1}{\tau_u} + \dfrac{1}{\tau_d}} = \frac{\tau}{\tau_d}\quad, \tag{11}$$

while the probability $p_d$ of down level is $\dfrac{\tau}{\tau_u}$, so

$$p_u p_d = p_u(1 - p_u) = \frac{\tau^2}{\tau_u \tau_d}\quad. \tag{12}$$

The power spectrum $S(\omega)$ of a single RTS is Lorentzian [2], that is

$$S(\omega) = (u_1 - u_2)^2 \, p_u \, (1 - p_u) \frac{4\tau}{1 + \omega^2 \tau^2} \qquad , \qquad (13)$$

where $p_u$ denotes the probability of finding the level $u_1$ (up level), $1-p_u$ the probability of finding the level $u_2$ (down level), $u_1 - u_2 = u$ and $(u_1 - u_2)^2 \, p_u \, (1 - p_u)$ the variance of the process.

The interevent time $\tau_k$ between two successive rising (or falling) edges fluctuates around an average time

$$\tau = \tau_u + \tau_d \quad . \qquad (14)$$

We can suppose Gaussian the distribution of $\tau_k$, that is

$$t_k - t_{k-1} = \tau_k = \tau + \sigma \, \varepsilon_k \qquad , \qquad (15)$$

where $t_k$ is the occurrence time of the rising (or falling) edges of the RTS, $\varepsilon_k$ is a random variable normally distributed around the zero with unity variance and $\sigma$ is the standard deviation of this fluctuation.

As a consequence of Eq.(15), the interval $\tau_k$ undergoes a Brownian increase or decrease.

This representation of the interevent time of an RTS is consistent with the Lorentzian power spectrum which appears at low frequency if $f < \Delta t_p$ with $\Delta t_p$, being the characteristic pulse length, no matter the shape of pulses [10].

## 5. Superposition of single-sided RTSs

When many random-phased single-sided RTSs, of same amplitude and relaxation time $\tau$, are summed, they are cross-correlate by their coincidences.

Differently from the rectangular signals, a coincidence among various RTSs phases the signals only for a limited time (*phasing time*), according the value of *c*. Indeed, the Brownian spread of the occurrence time of the RTSs is statistically symmetric, so the emersion of the pulses of coincidence degrades in the time (*phasing time*) and the temporal distance between the maxima of whatever two pulses of coincidence, within the same phasing time, will be averagely $n\tau$, with $n$ an integer number.

Therefore, as for the rectangular signals, the structure of the originating RTSs reappears on the shape and on the average periodicity of the emerging pulses in the resulting process $U$. This periodicity persists within the phasing time which can be intended as a relaxation time $\tau_{ph}$ (or memory). A part amplitude and shape of the pulses of coincidence, various relaxation times can appear in the resulting process, which can be decomposed in a set of independent Lorentzians of different $\tau_{ph}$ and variance, which mainly influence the high frequency power spectrum [10].

This results evident in Fig.2, where we can see the decomposition of a simple random sequence of phasing times of a resulting process, in three Lorentzians with different $\tau_{ph}$ . The lack of coincidences in these three Lorentzians is obvious, while their variance is decreasing with $\tau_{ph}$ , according to the rarity of the fluctuations in up level.

In conclusion, in the superposition of similar RTSs, the pulses due to the coincidences of the originating signals generate a process which can be interpreted by a sequence of phasing times. The pulses within a phasing time have a periodicity self-adjusting to an average value $\tau$. As a consequence, the resulting process can be interpreted by a sum of independent Lorentzians, with different relaxation times $\tau_{ph}$ and variances. A single Lorentzian, with the relaxation time of the originating RTSs, appears in addition to this set of independent Lorentzians. We shall see that this set of Lorentzians generates $1/f$ noise, while the presence of a Lorentzian, in addition to $1/f$ noise, is ascertained in various experiments [11][12][13].

As concerns the shape of the pulses, it reveals a self-organization process, such as the generation of stepped symmetrical triangles (see Fig.3A).

### 5.1. *The fluctuation of the coincidences*

Consider the number $c$ of *coincidences* which result in the superposition of $N \gg 1$ single-sided RTSs, with the same amplitude $u$ and relaxation time $\tau$ .

Call $P(c)$ the probability of $c$ coincidences, in the sampling time $\Delta T$, and $\Delta c$ the corresponding double-sided fluctuation of $c$ from the mean $\langle c \rangle$ and $P(\Delta c)$ the probability of fluctuation $\Delta c$.

If $U$ represents the summation of all $u_i$ signals, we call $\Delta U$ the double-sided fluctuation of $U$ from the mean $\langle U \rangle$, being $U=cu$ and $\Delta U=\Delta c\ u$.

The particularity of the amplitudes $\Delta U$ is their *discrete value,* for they can be reached only through the eventuality of an increment $\Delta c = \Delta U/u$.

When the number of the originating RTSs is high, and the probability of up level is not too small (or too high), the distribution of the amplitudes $\Delta c$ approaches a Gaussian.

Then the fluctuations $\Delta c$ will follow the relationship

$$P(\Delta c) \propto e^{-\frac{\Delta c^2}{2\sigma_{\Delta c}^2}} \qquad . \qquad\qquad (17)$$

The same distribution is worth for the fluctuations $\Delta U$, thanks to the originating RTSs of the same amplitude. Thus the probability $P(\Delta U)$ of the fluctuation $\Delta U$ results

$$P(\Delta U) \propto e^{-\frac{\Delta U^2}{2\sigma_{\Delta U}^2}} \qquad , \qquad\qquad (18)$$

where $\sigma_{\Delta U}{}^2$ represents the variance of this distribution.

The memory of the originating RTSs does not compromise the final distribution of the coincidences, for it depends only from the probability $p$ and from $N$. In any case this memory is transmitted to the resulting process $\Delta U$ on the shape and the duration of its pulses.

### 5.2.  *Self-similar average periodic structure*

Fig.3A represents the resulting process $U$ due the superposition of five similar single-sided originating RTSs of amplitude $u$. The process $U$ has the following properties, when the number of the originating RTSs is high:

- The coincidences self-organize symmetrical graded pulses of triangular shape, which amplitudes change for the quantity $u$, or multiple of $u$.
- The fluctuations $\Delta U$ from the mean $<U>$ obey, in general, to a Gaussian distribution and, as a consequence, also the maxima of the pulses $\Delta U$ will obey to the same distribution.
- The time series of the maxima of the pulses $\Delta U$ of the same amplitude is phased on the time series of the maxima of immediately minor amplitude and so on.
- If the maxima of the pulses $\Delta U$ obey to a Gaussian distribution, also the corresponding fluctuations $\Delta c$ obey to the same distribution (Sec.5.1).

In Fig.4 it is represented, for sake of simplicity and necessity of scale, the self-similar periodic average pattern which obeys to the distribution

$$P(\Delta c) \propto 2^{-\Delta c} \quad , \tag{19}$$

where, in comparison with the distribution of Eq.(17), 2 is put in the place of $e$, $\Delta c$ in the place of $\Delta c^2$, and $2\,\sigma_{\Delta U}^2 = 1$. So, for $\Delta c = 1, 2, 3, 4, \ldots$ the average frequency of the fluctuations scales averagely according to the series 1/2, 1/4, 1/8, 1/16, …, otherwise using $\Delta c^2$ the series would have been 1/2, 1/4, 1/16, 1/256, … .

The pattern of Fig.4 recalls the so-called *Voss-McCartney algorithm* generating $1/f$ noise [15] and it is reminiscent of an one-dimensional Sierpinsky signal whose structure can generate $1/f^\alpha$ spectra [16].

In conclusion, the self-similar structure of the fluctuations $\Delta U$ can be intended only averagely. Indeed, a structure merely periodic would be generate a discrete power spectrum and it is meaning that *Voss-McCartney algorithm* can be improved by adding a white noise source [15].

## 6. 1/*f* power spectrum

Various versions of the model of the superposition of many similar single-sided RTSs can be adopted to demonstrate the presence of 1/*f* power spectrum on the resulting process. Indeed, facing a general version (sum of independent Lorentzians) other versions are possible, based on particular statistical aspect of the resulting process.

### 6.1. *Sum of independent Lorentzians*

It is known that a weighed summation of Lorentzians, with a distribution of relaxation times $\tau$ in the range $\tau_2^{-1} << \tau^{-1} << \tau_1^{-1}$, yields a 1/*f* power spectrum. Indeed, with a weight proportional to $1/\tau$ and operating in the frequency range $\tau_2^{-1} << 2\pi f << \tau_1^{-1}$, we have [2] [6]

$$S(f) \propto \int_{\tau_2}^{\tau_1} \left(\frac{1}{\tau}\right) \frac{\tau}{1 + (2\pi f)^2 \tau^2} d\tau \propto \frac{1}{f} \quad . \tag{20}$$

Although the Lorentzians of Fig.3C generate the process of Fig.3A, the power spectrum of the resulting process cannot be built by a simple weighted summation of these Lorentzians (using their variance), because the resulting pulses are obtained by coincidences and in this sense the cross-correlation term cannot be negliged. Besides, the variance of these Lorentzians is not inversely proportional to $\tau$, but rather directly proportional.

Therefore, the appearance of 1/*f* power spectrum requires adjunctive characteristics in the resulting process, such as the emersion of coincidences which phase the fluctuations of the process itself (Sec.5). Hence, looking to the time series of the phasing times (or relaxation times) of the resulting process, we can decompose this sequence in various independent Lorentzians with a distribution of relaxation times (see example in Fig.2). The variance of these Lorentzians is just inversely proportional to $\tau$, due to the rarity of the fluctuations with the increase of the relaxation time.

Supposing a continuous distribution of relaxation times, the power spectrum $S(f)_U$ has to be averaged over this distribution with a function weight $p(\tau)$ $d\tau \propto 1/\tau$ $d\tau$, which includes the contribution to the variance of processes whose relaxation time lies in the interval from $\tau$ to $\tau + d\tau$ .

The cross-correlations terms do not appear in the superposition of these Lorentzians, because these terms simply do not exist, being the Lorentzians derived by decomposing a resulting process and not vice versa.

### 6.2. Point process model

This model can be considered as a particular case of the sequence of phasing times of the resulting process due to the coincidences of many similar single-sided RTSs, mentioned in Sec.5. Indeed, this sequence can appear without interruptions and in this circumstance, the occurrence times $t_k$, of the maxima of the resulting pulses, can be described by the recurrent equations [11][16]

$$
\begin{aligned}
t_k &= t_{k-1} + \tau_k \\
\tau_k &= \tau_{k-1} - \gamma(\tau_{k-1} - \tau) + \sigma \varepsilon_k
\end{aligned} \qquad , \qquad (21)
$$

where $\gamma$ is a number $\ll 1$ representing the average strength which approaches $\tau_k$ to the average value $\tau$, while $\{\varepsilon_k\}$ denotes the sequence of uncorrelated normally distributed random variables with zero expectation and unit variance (white noise source) and $\sigma$ is the standard deviation of the white noise.

Eqs. (21) exhibit an autoregressive process with very small damping, in the sense that when $\tau_{k-1}$ is different from $\tau$ it is adjusted to be more (normally distributed) near to $\tau$.

After some algebra, and in the frequency range $f_1 < f < f_2, f_\tau$, being $f_1 = \gamma^{3/2}/\pi\sigma, f_2 = 2\gamma^{1/2}/\pi\sigma, f_\tau = (2\pi\tau)^{-1}$, a $1/f$ spectrum is obtained from the series of the times $t_k$.

This model, based on a self-adjusting average periodicity of occurrence times $t_k$, depends only on statistics and correlations of occurrence times $t_k$, because the shape of pulses mainly influence the high frequency power spectral density [11][16].

We can also to note that processes as a *pure periodic*, or *Poisson*, can be easily excluded, as also a *perturbed periodic process*, expressed by the recurrent equation $\tau_k = \tau + \sigma \varepsilon_k$, due to the *infinite Brownian increase or decrease* of the intervals $\tau_k$.

### 6.3. Gaussian model

A Gaussian noise not necessarily has a $1/f$ power spectrum, but not even it excludes its presence. Gaussian noise concurs to $1/f$ power spectrum when the resulting process assumes the aspect of a self-similar periodic average signal (Sec.5.2). In this case, we can calculate the power $\Delta U^2$ from Eq.(18), that is

$$
\Delta U^2 = 2\sigma_{\Delta U}^2 \left[ \ln \frac{1}{P(\Delta U)} + const \right]. \qquad (22)
$$

where $\Delta U^2$ represents *the average power of the events which have the same probability $P(\Delta U)$* in the measurement time $T$. Indeed, if each measurement requires the time $\Delta T$, various values of power can appear with the same probability $P(\Delta U)$, on $T/\Delta T$ alternatives. On the other hand, the *law of large numbers* allows us to introduce *rates* (average number of events in time unity) in the place of probabilities, thereby enabling us to substitute the probability $P(\Delta U)$ with the *rate* $r_{\Delta U} = P(\Delta U)/\Delta T$.

Therefore, Eq.(22) becomes

$$\Delta U^2 = 2\sigma^2_{\Delta U}(\ln\frac{1}{r_{\Delta U}} + const) \quad , \qquad (23)$$

where $\Delta U^2$ denotes *the average power of all the fluctuations of rate $r_{\Delta U}$*.

If the average structure of process $\Delta U$ tends to become periodic on any frequency scale (self-similar structure), the various *rates $r_{\Delta U}$* tend to become discrete, assuming the meaning of frequencies $f_{\Delta U}$. So, $\Delta U^2$ denotes also *the average power of all the fluctuations of frequency $f_{\Delta U}$* or, in other words, *of all the fluctuations repeated with a period $1/f_{\Delta U}$*.

$$\Delta U^2 = 2\sigma^2_{\Delta U}(\ln\frac{1}{f_{\Delta U}} + const) \quad . \qquad (24)$$

Now, the periodic average process $\Delta U$, of fundamental frequency $f_{\Delta U}$, has spectral components in the frequency range $f_{\Delta U}, f_{max}$ (the half of the sampling frequency) and its average power is just the summation of the powers of its harmonics [17].

Hence, supposing a continuous distribution of the spectral components, we can introduce the power spectrum $S(f)_{\Delta U}$, so that the power increment, in the frequency interval $f_{\Delta U_2} < f_{\Delta U_1}$ will be

$$\int_{f_{\Delta U_1}}^{f_{\Delta U_2}} S(f)_{\Delta U}\, df = 2\sigma^2_{\Delta U}\ln\frac{f_{\Delta U_1}}{f_{\Delta U_2}} \quad , \qquad (25)$$

that is

$$S(f)_{\Delta U} = \frac{2\sigma^2_{\Delta U}}{f} \quad . \qquad (26)$$

In conclusion, tanks to a periodic average self-similar structure, a Gaussian process can generate $1/f$ power spectrum.

Vice versa, the $1/f$ power spectrum of a self-similar and periodic process $\Delta U$, conducts to a Gaussian distribution. Indeed, the average power of the fluctuation $\Delta U$ in the frequency range $f_{max}, f_{\Delta U}$, will be

$$\Delta U^2 = \int_{f_{\Delta U}}^{f_{max}} \frac{const}{f}\, df \propto \ln\frac{f_{max}}{f_{\Delta U}} \qquad (27)$$

that is

$$f_{\Delta U} = f_{max} \; e^{-\frac{\Delta U^2}{const}} \;,$$

(28)

where $\Delta U^2$ represents the average power of the process $\Delta U$ in the period $1/f_{\Delta U}$.

The Gaussian distribution of $1/f$ noise is found in measurements performed by R.F. Voss on different solid state devices [18]. In particular R.F. Voss tested the correlation between Gaussian behaviour and $1/f$ noise observed in a sufficiently pure form.

### 6.4. *"Running" sand-pile model*

It is not difficult to translate the generation of coincidences, due to the summation of many single-sided RTSs, in a version of the SOC (self-organized criticality) model [19]. Indeed, we can adopt the continuous one-dimensional version of the SOC, that is the so-called *"running" sand-pile model* [20], also described with new analyses and signatures in the reference [21].

The basic model consists of a one-dimensional lattice of $L$ cells. The boundary of the lattice at $n = 0$ is closed, while at $n = L$ is open. Each cell contains an integer number of sand grains which represents the height of the cell.

To every cell $n$ is associated the local gradient $z_n$ representing the difference in height between two neighbouring cells.

The transport process is initiated by randomly depositing sand grains into the system. At each time step, there is a probability $p$ of depositing $N_0$ grains of sand to each site. The system can started from uniform or random conditions.

The configuration $z_n$ is *simultaneously* updated according to the following evolution rules

$$z_n(t+1) = z_n(t) - N_f$$
$$z_{n\pm1}(t+1) = z_{n\pm1}(t+1) + N_f$$

(29)

if and only if $z_n(t+1) \geq N_f$.

After some time, the system reaches a steady state in which the input is on average balanced by the drainage at the open end.

The local gradient $z_n$ represents, in our case, an RTS, random switching from the value zero to one, when $z_n \geq N_f$.

The activity of the system is the monitored by the total number $c(t)$ of the unstable cells, or *flips* (the coincidences in our case), at each time step. The chain reaction of updates is referred to as an *avalanche*.

The ingredient of the memory appears in the system when the sites are not updated in consecutive time steps.

According to references [20][21], $1/f$ noise on $c(t)$ may arise as a consequence of a suitable deposition rate of grains of sand and of a system size ($L$). This happens when the set $z_n$ mimes the set of single-sided RTS with the same amplitude, probability in up level and relaxation time. In the case of ref.[20], it is interesting to note that the displayed waveform of the fluctuations of $c(t)$ is just similar to that drawn in Fig.1A. In conclusion, the running sandpile model, as concerns the generation of $1/f$ noise cannot be seen as a general physical model for $1/f$ noise, but as physical mechanism useful to generate a set of similar single-sided RTSs.

## 7. Conclusions

In this paper it has been demonstrated that $1/f$ noise is produced by a superposition of many single-sided RTSs with the same amplitude, probability and relaxation time. This result is possible thanks to:

- The presence of the coincidences, or cross-correlations, among the various originating RTSs.
- The self-adjusting periodicity of the fluctuations with various phasing times (or relaxation times).
- The decomposition of the sequence of these phasing times by a set of Lorentzians with a distribution relaxation times.

Particular versions of this model are:

- A resulting process with a sequence of phasing times without interruptions (*point process model*).
- A resulting process with a self-similar periodic average structure (*Gaussian model*).
- The generation of a superposition of similar single-sided RTSs by a "*running*" *sand-pile model*.

**Figure captions**

Fig. 1:  (B) Superposition of five random phased similar rectangular signals (see text for discussion).

Fig. 2:  Decomposition of a sequence of phasing times in three independent Lorentzians with different relaxation times (see text for discussion).

Fig. 3:  (A) Resulting process due to the superposition of the five RTSs (B). (C) Equivalent set of cross-correlated RTSs to achieve the resulting process (A) (see text for discussion).

Fig. 4:  Average pattern of fluctuations $\Delta c$ due to the superposition of many similar single-sided RTSs, according to the probability distribution $P(\Delta c) \propto 2^{-\Delta c}$ (see text for discussion).

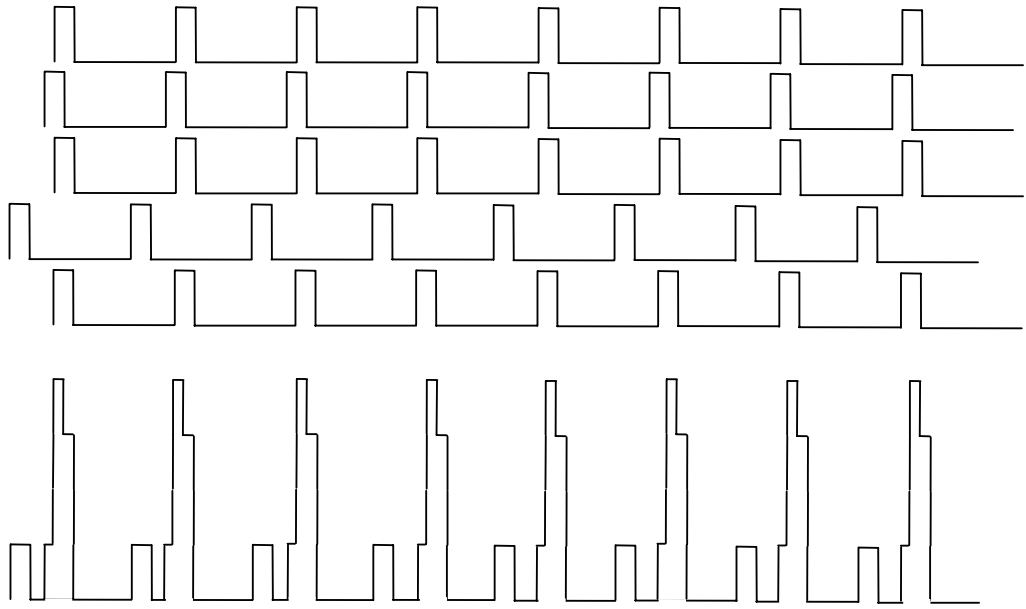

**Fig. 1**

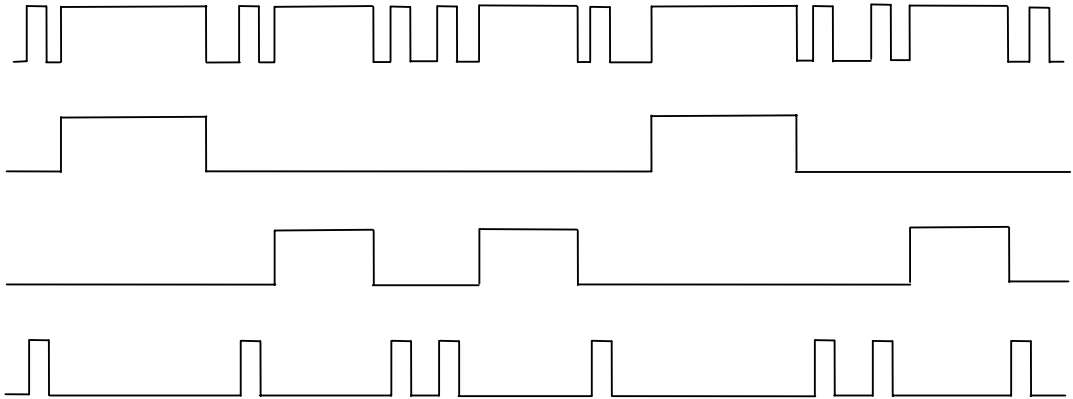

**Fig. 2**

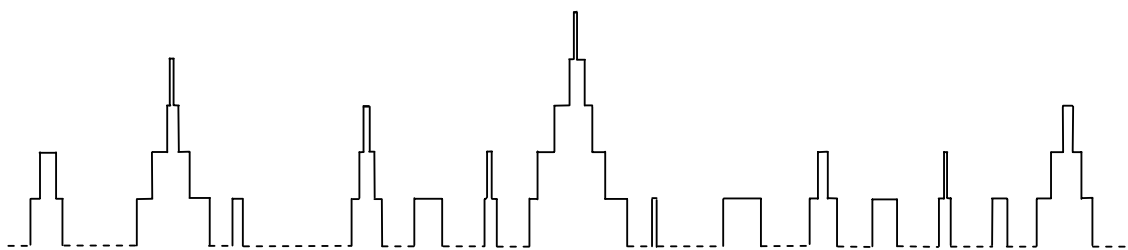

A

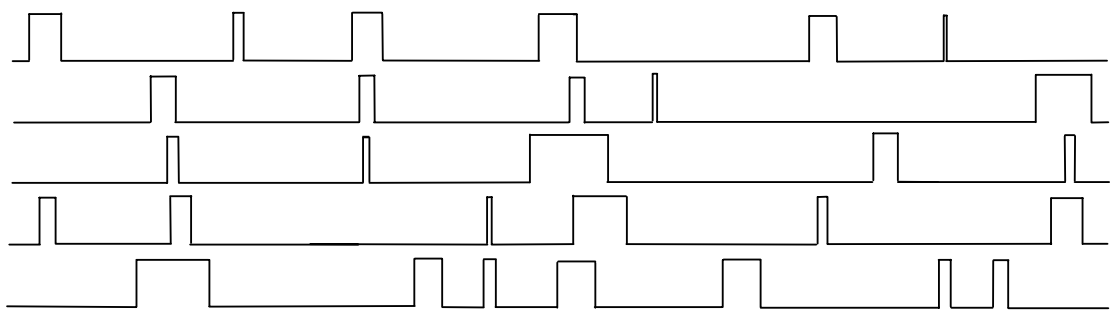

B

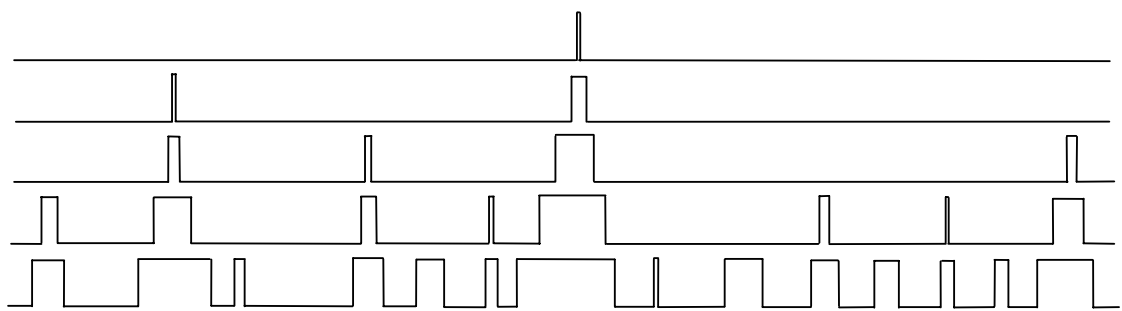

C

**Fig. 3**

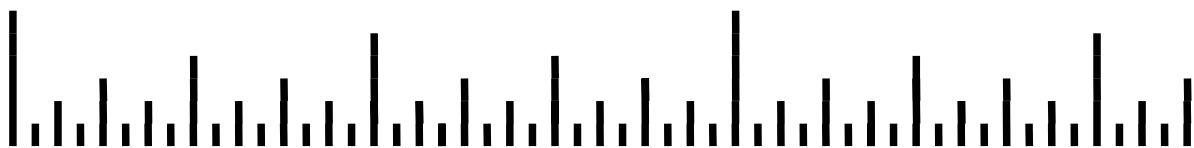

**Fig. 4**